\documentclass[12pt]{article}
\usepackage{amsthm,amssymb,amsmath}

\newtheorem*{Th}{Theorem}
\newtheorem*{pro}{Proposition}

\newtheorem{lem}{Lemma}

\newcommand{\dif}{\operatorname{d}}
\renewcommand{\t}{{\operatorname{t}}}

\newcommand{\bX}{{\boldsymbol X}}

\newcommand{\bx}{{\boldsymbol x}}
\newcommand{\be}{{\boldsymbol e}}
\newcommand{\bc}{{\boldsymbol c}}
\newcommand{\bu}{{\boldsymbol u}}

\newcommand{\bxi}{{\boldsymbol \xi}}
\newcommand{\bzeta}{{\boldsymbol \zeta}}

\newcommand{\R}{{\mathbb R}}

\newcommand{\D}{{\partial}}

\begin{document}

\title{Vectorial Ribaucour Transformations\\
for  the Lam\'e Equations}

\author{Q. P. Liu$^1$\thanks{On leave of absence from
Beijing Graduate School, CUMT, Beijing 100083, China}
\thanks{Supported by {\em Beca para estancias temporales
de doctores y tecn\'ologos extranjeros en
Espa\~na: SB95-A01722297}}
   $\,$ and Manuel Ma\~nas$^{1,2}$\thanks{Partially supported by CICYT:
 proyecto PB95--0401}\\
$^1$Departamento de F\'\i sica Te\'orica,\\ Universidad Complutense,\\
E28040-Madrid, Spain.\\
$^2$Departamento de Matem\'atica Aplicada y Estad\'\i stica,\\
Escuela Universitaria de Ingenier\'\i{}a T\'ecnica Areona\'utica,\\ 
Universidad Polit\'ecnica de Madrid,\\
E28040-Madrid, Spain.}
\date{}

\maketitle

\begin{abstract}
The vectorial extension of the Ribaucour transformation for the
Lam\'e equations of orthogonal conjugates nets in multidimensions
is given. We show that the composition of two vectorial Ribaucour
transformations with appropriate transformation data is again a
vectorial Ribaucour transformation, from which it follows the
permutability of the vectorial Ribaucour transformations. Finally,
as an example we apply  the vectorial Ribaucour transformation to
the Cartesian background.
\end{abstract}
\newpage


{\bf 1}. The connection between Soliton Theory and Differential
Geometry of Surfaces in Euclidean Space is well established. Many
systems considered in Geometry have been analyzed independently in
Soliton Theory, as examples we cite the Liouville and sine-Gordon
equations which characterize minimal and pseudo-spherical surfaces,
respectively. An important case is given by the the Darboux
equations for conjugate systems of coordinates that were solved 12
years ago in its matrix generalization, using the
$\bar\partial$--dressing, by Zakharov and Manakov \cite{zm}, and
further the Lam\'e equations for orthogonal conjugate nets were
solved only very recently \cite{z} by Zakharov imposing appropriate
constraints in the Marchenko integral equation associated with the
Darboux equations.

In this note we present a vectorial extension of  a transformation
that preserves the Lam\'e equations which is known as Ribaucour
transformation \cite{Ribaucour}. This vectorial extension can be
thought as the result of the iteration of the standard Ribaucour
transformation; i. e. sequences of Ribaucour transformations. The
expressions that we found are expressed in terms of multi-Grammian
type determinants, as in the fundamental transformation case.

The layout of this latter is as follows. In \S 2 we recall the
reader the Darboux system for conjugate nets and its vectorial
fundamental transformations, then in \S 3 we present the Lam\'e
equations for orthogonal conjugate nets and show how the vectorial
fundamental transformation reduces to the vectorial Ribaucour
transformation. Here we also prove that, given an orthonormal basis
of tangent vectors to the orthogonal conjugate coordinate lines,
the vectorial Ribaucour transformation preserves this character.
Next, in \S 4 we prove the permutability for the vectorial
Ribaucour transformation basing the discussion in a similar
existing result for the vectorial fundamental transformation.
Finally, in \S 5 we present an example: we dress the zero
background, specifically the Cartesian coordinates.

{\bf 2}. The Darboux equations
\begin{equation} \label{dar}
\frac{\partial\beta_{ij}}{\D u_k}=\beta_{ik}\beta_{kj},
\;\;
i,j,k=1,\dotsc, N,\;
\text{with $i,j,k$ different},\end{equation}
for the $N(N-1)$ functions
$\{\beta_{ij}\}_{\substack{i,j=1,\dotsc,N\\i\neq j}}$ of
$\bu:=(u_1,\dotsc,u_N)$, characterize $N$-dimen\-sional
submanifolds of $\R^D$, $N\leq D$, parametrized by conjugate
coordinate systems \cite{Darboux1,Eisenhart1}, and are the
compatibility conditions of the following linear system
\begin{equation} \label{X}
\frac{\partial \bX_j}{\D u_i} = \beta_{ji} \bX_i, \quad i,j=1,\dotsc,N,\quad
i\ne j,
\end{equation}
involving suitable $D$-dimensional vectors $\bX_i$, tangent to the
coordinate lines.
 The so called Lam\'e coefficients satisfy
\begin{equation} \label{H}
\frac{\partial H_j}{\D u_i} = \beta_{ij} H_i, \quad i,j=1,\dotsc,N,\quad
i\ne j,
\end{equation}
and the points of the surface $\bx$ can be found by means of
\begin{equation}\label{points}
\frac{\D \bx}{\D u_i}= \bX_i H_i,\quad i=1,\dotsc, N,
\end{equation}
which is equivalent to the more standard Laplace equation
\[
\frac{\D^2\bx}{\D u_i\D u_j}=\frac{\D \ln H_i}{\D u_j}\frac{\D \bx}{\D u_i}
+\frac{\D \ln H_j}{\D u_i}\frac{\D \bx}{\D u_j},
\quad i,j=1,\dotsc,N,\;\; i\neq j.
\]

The fundamental transformation for the Darboux system was
introduced by \cite{jonas,Eisenhart3}, see also
\cite{Eisenhart2,ks}, and its vectorial extension was given, in a
discrete framework in \cite{mds,dsm}. It requires the introduction
of a potential in the following manner: given vector solutions
$\bxi_i\in V$ and $\bzeta_i^*\in W^*$ of \eqref{X} and \eqref{H},
$i=1,\dotsc,N$,
 respectively, where
$V,W$ are linear spaces and $W^*$ is the dual space of $W$,
one can define a potential matrix
$\Omega(\bxi,\bzeta^*):W\to V$ through the equations
\begin{equation}\label{potential}
\frac{\D   \Omega(\bxi,\bzeta^*)}{\D u_i}=\bxi_i\otimes\bzeta^*_i.
\end{equation}
We give here the continuous version of the
vectorial fundamental transformation for quadrilateral lattices
\cite{mds,dsm}.

\newtheorem*{VF}{Vectorial Fundamental Transformation}

\begin{VF}
Given solutions $\bxi_i\in V$ and $\bxi_i^*\in V^*$ of
\eqref{X} and \eqref{H}, $i=1,\dotsc,N$,
 respectively,
new rotation coefficients $\hat\beta_{ij}$, tangent vectors
$\hat{\bX}_i$, Lam\'e coefficents $\hat H$ and points of the
surface $\hat{\bx}$ are given by
\begin{equation}\label{vecfun}
\begin{aligned}
\hat\beta_{ij}&=\beta_{ij}-
\langle\bxi^*_j, \Omega(\bxi,\bxi^*)^{-1}\bxi_i\rangle,\\
\hat{\bX}_i&=\bX_i-\Omega(\bX,\bxi^*)\Omega(\bxi,\bxi^*)^{-1}\bxi_i,\\
\hat H_i&=H_i-\bxi_i^*\Omega(\bxi,\bxi^*)^{-1}\Omega(\bxi,H),\\
\hat{\bx}&=\bx-\Omega(\bX,\bxi^*)\Omega(\bxi,\bxi^*)^{-1}\Omega(\bxi,H).
\end{aligned}
\end{equation}
\end{VF}
\noindent
Here we are assuming that $ \Omega(\bxi,\bxi^*)$ is invertible.
We shall refer to this transformation as vectorial fundamental
transformation with transformation data $(V,\bxi_i,\bxi_i^*)$.

{\bf 3}.  The Lam\'e equations describe $N$-dimensional conjugate
orthogonal systems of coordinates \cite{lame,Darboux2,tsarev}:
\begin{align}
&\frac{\partial\beta_{ij}}{\D u_k}-\beta_{ik}\beta_{kj}=0,\;\;
i,j,k=1,\dotsc, N,\;
\text{with $i,j,k$ different},\label{lame1}\\
&\frac{\partial\beta_{ij}}{\D
u_i}+
\frac{\partial\beta_{ji}}{\D u_j}+
\sum_{\substack{k=1,\dotsc,N\\ k\neq i,j}}
\beta_{ki}\beta_{kj}=0,\quad i,j=1,\dotsc,N,\;
i\neq j.\label{lame2}
\end{align}

An important observation, that in the scalar case appears in \cite{schief}, is:
\begin{lem}
Given a solution $\bxi_i\in V$ of \eqref{X} then
\begin{equation}\label{*}
\bxi^*_i:=\Big(\frac{\D\bxi_i}{\D u_i}+
\sum_{\substack{k=1,\dotsc,N\\ k\neq i}}\bxi_k \beta_{ki}\Big)^{\t},
\end{equation}
where $^\t$ means transpose, is a $V^*$-valued solution of \eqref{H}
if and only if \eqref{lame2} holds.
\end{lem}
\begin{proof}
Just notice that from \eqref{dar} and \eqref{X} it follows that
 \[
\frac{\D\bxi^*_j}{\D u_i}=
\beta_{ij}\bxi_i^*+\Big(\frac{\partial\beta_{ij}}{\D
u_i}+\frac{\partial\beta_{ji}}{\D u_j}+
\sum_{\substack{k=1,\dotsc,N\\ k\neq i,j}}
\beta_{ki}\beta_{kj}\Big)\bxi_i^\t.
\]
\end{proof}

A second observation is that,
\begin{lem}
 Given $\beta$'s solving the Lam\'e
equations \eqref{lame1} and \eqref{lame2}, $\bxi_i\in V$ and
$\bzeta_i\in W$ solutions of
\eqref{X} and $\bxi^*_i$ and $\bzeta^*_i$ as prescribed in \eqref{*},
$i=1,\dotsc,N$, then:
\[
\frac{\D}{\D u_i}\Big(\Omega(\bxi,\bzeta^*)+\Omega(\bzeta,\bxi^*)^\t-
\sum_{k=1,\dotsc,N}\bxi_k\otimes\bzeta_k^\t\Big)=0,\quad i=1,\dotsc,N,
\]
\end{lem}
\begin{proof}
Using \eqref{potential} and the definition \eqref{*} we have
\begin{multline*}
\frac{\D}{\D u_i}(\Omega(\bxi,\bzeta^*)+\Omega(\bzeta,\bxi^*)^\t)
=\bxi_i\otimes\Big(\frac{\D\bzeta_i}{\D u_i}+
\sum_{\substack{k=1,\dotsc,N\\ k\neq i}}\bzeta_k \beta_{ki}\Big)^{\t}\\+
\Big(\frac{\D\bxi_i}{\D u_i}+
\sum_{\substack{k=1,\dotsc,N\\ k\neq i}}\bxi_k \beta_{ki}\Big)
\otimes\bzeta_i^{\t},
\end{multline*}
and recalling that $\bxi_i\beta_{ki}=(\D /\D u_i)\bxi_k$ and
$\bzeta_i\beta_{ki}=(\D /\D u_i)\bzeta_k$ we get the statement.
\end{proof}

Therefore, as $\Omega(\bxi,\bzeta^*)$ and $\Omega(\bzeta,\bxi^*)$
are defined by \eqref{potential} up to additive constant matrices,
the previous lemma is telling us that we can take those constants
such that
\begin{equation*}
\Omega(\bxi,\bzeta^*)+\Omega(\bzeta,\bxi^*)^\t=
\sum_{k=1,\dotsc,N}\bxi_k\otimes\bzeta_k^\t.
\end{equation*}

We now can prove the following lemma:
\begin{lem}
Suppose given a solution $\beta_{ij}$ of the Lam\'e equations
\eqref{lame1} and \eqref{lame2},  $\bxi_i\in V$ and $\bzeta_i\in W$
solving \eqref{X} and  $\bxi_i^*$ and $\bzeta_i^*$ as prescribed in
\eqref{*}. Then,
if
\begin{equation}\label{constraints}
\begin{aligned}
\Omega(\bxi,\bzeta^*)+\Omega(\bzeta,\bxi^*)^\t&=
\sum_{k=1,\dotsc,N}\bxi_k\otimes\bzeta_k^\t,\\
\Omega(\bxi,\bxi^*)+\Omega(\bxi,\bxi^*)^\t&=
\sum_{k=1,\dotsc,N}\bxi_k\otimes\bxi_k^\t
\end{aligned}
\end{equation}
 the vectorial fundamental transformation \eqref{vecfun}:
\begin{align*}
\hat\beta_{ij}&=\beta_{ij}-
\langle\bxi^*_j, \Omega(\bxi,\bxi^*)^{-1}\bxi_i\rangle,\\
\hat{\bzeta}_i&=\bzeta_i-\Omega(\bzeta,\bxi^*)\Omega(\bxi,\bxi^*)^{-1}\bxi_i,\\
\hat\bzeta_i^*&=\bzeta_i^*-\bxi_i^*\Omega(\bxi,\bxi^*)^{-1}\Omega(\bxi,\bzeta^*),
\end{align*}
is such that
\[
\hat\bzeta^*_i:=\Big(\frac{\D\hat\bzeta_i}{\D u_i}+
\sum_{\substack{k=1,\dotsc,N\\ k\neq i}}\hat\bzeta_k \hat\beta_{ki}\Big)^{\t}.
\]
\end{lem}
\begin{proof}
Using \eqref{potential}, \eqref{vecfun} and \eqref{*} we find that
\begin{multline*}
\frac{\D\hat\bzeta_i}{\D u_i}+
\sum_{\substack{k=1,\dotsc,N\\ k\neq i}}\hat\bzeta_k \hat\beta_{ki}=
(\bzeta_i^*)^\t-\Omega(\bzeta,\bxi^*)\Omega(\bxi,\bxi^*)^{-1}(\bxi_i^*)^\t
\\ -\sum_{k=1,\dotsc,N}\langle\bxi_i^*,\Omega(\bxi,\bxi^*)^{-1}\bxi_k\rangle
(\bzeta_k-\Omega(\bzeta,\bxi^*)\Omega(\bxi,\bxi^*)^{-1}\bxi_k),
\end{multline*}
that together with the identity
\[
\langle\bxi_i^*,\Omega(\bxi,\bxi^*)^{-1}\bxi_k\rangle=
\bxi_k^\t(\Omega(\bxi,\bxi^*)^{-1})^\t(\bxi_i^*)^\t,
\]
implies
\begin{multline*}
\frac{\D\hat\bzeta_i}{\D u_i}+
\sum_{\substack{k=1,\dotsc,N\\ k\neq i}}\hat\bzeta_k \hat\beta_{ki}=
(\bzeta_i^*)^\t-\Big[\Omega(\bzeta,\bxi^*)\Omega(\bxi,\bxi^*)^{-1}
\\+\sum_{k=1,\dotsc,N}
\Big(\bzeta_k-\Omega(\bzeta,\bxi^*)\Omega(\bxi,\bxi^*)^{-1}\bxi_k\Big)\otimes
\bxi_k^\t(\Omega(\bxi,\bxi^*)^{-1})^\t\Big](\bxi_i^*)^\t.
\end{multline*}
Now, the constraints \eqref{constraints} applied to the above
expression gives
\[
\frac{\D\hat\bzeta_i}{\D u_i}+
\sum_{\substack{k=1,\dotsc,N\\ k\neq i}}\hat\bzeta_k \hat\beta_{ki}=
(\bzeta_i^*)^\t-\Omega(\bxi,\bzeta^*)^\t
(\Omega(\bxi,\bxi^*)^{-1})^\t(\bxi_i^*)^\t,
\]
that when transposed gives the desired equality.
\end{proof}

With these  lemmas avaliable we are able to state the main theorem
of this letter:
\begin{Th}
The vectorial fundamental transformation \eqref{vecfun} when
applied to a solution of the Lam\'e equation preserves the
orthogonal character of the conjugate net whenever the
transformation data ($V$, $\bxi_i$, $\bxi_i^*$) satisfy
\begin{align*}
&(\bxi_i^*)^\t=\frac{\D\bxi_i}{\D u_i}+
\sum_{\substack{k=1,\dotsc,N\\ k\neq i}}\bxi_k \beta_{ki},\\
&\Omega(\bxi,\bxi^*)+\Omega(\bxi,\bxi^*)^\t=
\sum_{k=1,\dotsc,N}\bxi_k\otimes\bxi_k^\t.
\end{align*}
\end{Th}
\begin{proof}
Lemma 3 together with Lemma 1 imply that the new $\hat\beta$ are a
solution of the Lam\'e equations \eqref{lame1} and \eqref{lame2}.
\end{proof}

A vectorial fundamental transformation with data
$(V,\bxi_i,\bxi_i^*)$ as in the
Theorem will be referred as a vectorial Ribaucour
transformation with data $(V,\bxi_i)$. In the scalar case the
vectorial Ribaucour transformation reduces to the Ribaucour
transformation \cite{Ribaucour,Darboux2,Eisenhart2,tsarev}.

The Lam\'e equations \eqref{lame1} and \eqref{lame2} are the
compatibility conditions for
\begin{align}
\notag\frac{\D\bX_j}{\D u_i}&=\beta_{ji}\bX_i,\quad i,j=1,\dotsc,N\;i\neq j,\\
\label{2}\frac{\D\bX_i}{\D u_i}&=-\sum_{\substack{k=1,\dotsc,N\\ k\neq i}}\bX_k\beta_{ki}.
\end{align}
These conditions are equivalent to the fact that the independent
tangent vectors $\{\bX_i(\bu)\}_{i=1,\dotsc,N}$ form an orthonormal
basis for all $\bu$ if they do for a particular value $\bu_0$; i.
e. $\bX_i^\t\bX_j=\delta_{ij}$. We now show that the vectorial
Ribaucour transformation preserves this orthonormal character for
the transformed basis. Indeed, \eqref{2} together with
\eqref{*} implies $\bX_i^*=0$ and the vectorial fundamental
transformation gives $\hat\bX_i^*=0$ if $\Omega(\bxi,0)$, which is
an arbitrary constant matrix, is taken as zero. Hence, Lemma 3
implies that
\[
\frac{\D\hat\bX_i}{\D u_i}=
-\sum_{\substack{k=1,\dotsc,N\\ k\neq i}}\hat\bX_k\hat\beta_{ki},
\]
and recalling that
\[
\frac{\D\hat\bX_j}{\D u_i}=\hat\beta_{ji}\hat\bX_i,\quad i,j=1,\dotsc,N\;i\neq j,
\]
is satisfied, we find out that the new tangent vectors
$\{\hat\bX_i(\bu)\}_{i=1,\dotsc,N}$ form an orthonormal basis for
all $\bu$ if they do for some value of $\bu=\bu_0$. Indeed, by
choosing $\Omega(\bX,\bxi^*)=0$ one gets
$\bX_i(\bu_0)=\hat\bX_i(\bu_0)$; i. e. the initial basis and the
transformed one coincide at that point.

Notice that the above results constitute an alternative proof of
the Theorem.

{\bf 4}. In \cite{dsm} it was proven a permutability theorem for
the vectorial fundamental transformation of quadrilateral
lattices, here we give its continous limit to conjugate nets:

\newtheorem*{Per}{Permutability of Vectorial Fundamental Transformations}

\begin{Per}
The vectorial fundamental transformation with transformation data
\[
\bigg(V_1\oplus V_2,
\Big(\begin{smallmatrix} \bxi_{i,(1)}\\ \bxi_{i,(2)}\end{smallmatrix}\Big),
(\bxi^*_{i,(1)},\bxi^*_{i,(2)})\bigg)
\]
coincides with the following
composition of vectorial fundamental transformations:
\begin{enumerate}
\item First transform with data
\[
(V_2,\bxi_{i,(2)},\bxi_{i,(2)}),
\]
 and denote the transformation by $^\prime$.
\item On the result of this transformation apply a second one with data
 \[
(V_1, \bxi_{i,(1)}^\prime, \bxi_{i,(1)}^\prime).
\]
\end{enumerate}
\end{Per}

\noindent
Therefore, the composition of two vectorial fundamental
 transformations yields, independently of the order, a
new vectorial fundamental transformation; hence the permutability
character of these transformations. Moreover, from this result it
also follows that the vectorial fundamental transformation is just
a superposition of a number of fundamental transformations.

One can easily conclude that this result can be extended to
the vectorial Ribaucour transformation for orthogonal conjugate
nets.

\begin{pro}
The vectorial Ribaucour transformation with transformation data
\[
\bigg(V_1\oplus V_2,
\Big(\begin{smallmatrix} \bxi_{i,(1)}\\ \bxi_{i,(2)}\end{smallmatrix}\Big)\bigg),
\]
 as prescribed in our Theorem,  coincides with the following
composition of vectorial Ribaucour transformations:
\begin{enumerate}
\item First transform with data
\[
(V_2,\bxi_{i,(2)}),
\]
 and denote the transformation by $^\prime$.
\item On the result of this transformation apply a second one with data
   \[
(V_1, \bxi_{i,(1)}^\prime).
\]
\end{enumerate}
\end{pro}

\begin{proof}
Because the transformation data follows the prescription of our Theorem
they must satisfy
\begin{align*}
&(\bxi_{i,(s)}^*)^\t=\frac{\D\bxi_{i,(s)}}{\D u_i}+
\sum_{\substack{k=1,\dotsc,N\\ k\neq i}}\bxi_{k,(s)} \beta_{ki},\quad s=1,2\\
&\Omega(\bxi_{(s)},\bxi_{(s)}^*)+\Omega(\bxi_{(s)},\bxi_{(s)}^*)^\t=
\sum_{k=1,\dotsc,N}\bxi_{k,(s)}\otimes\bxi_{k,(s)}^\t,\quad s=1,2\\
&\Omega(\bxi_{(1)},\bxi_{(2)}^*)+\Omega(\bxi_{(2)},\bxi_{(1)}^*)^\t=
\sum_{k=1,\dotsc,N}\bxi_{k,(1)}\otimes\bxi_{k,(2)}^\t.
\end{align*}
Thus, we see that the first vectorial fundamental transformation is
a vectorial Ribaucour transformation with data $(V,\bxi_{i,(2)})$.
Now, applying Lemma 3, we see that the vectorial fundamental
transformation of point 2. is also a vectorial Ribaucour
transformation.
\end{proof}

{\bf 5}. For the zero background $\beta_{ij}=0$ we have that the
solutions of \eqref{X} are any set of functions $\{\bxi_i\}_{i=1,\dotsc,N}$
of the form
\[
\bxi_i=\bxi_i(u_i)\in\R^M
\]
and for the adjoint we have
\[
\bxi_i^*=\frac{\dif\bxi_i^t}{\dif u_i}.
\]
We also have
\[
\Omega(\bxi,\bxi^*)(\bu)=\sum_{k=1,\dotsc, N}\Omega_i(u_i)
\]
 with
\begin{align*}
&
\Omega_i(u_i)=\int_{u_{i,0}}^{u_i}\dif u_i\,\bxi_i\otimes\frac{\dif
\bxi_i^\t}{\dif u_i}+\Omega_{i,0} ,\\
&\Omega_{i,0}+\Omega_{i,0}^\t=
(\bxi_i\otimes\bxi_i^\t)\big|_{u_{i,0}}.
\end{align*}
 In particular, the
Cartesian background has $\bX_i=\be_i$, $\{\be_i\}_{i=1,\dotsc,N}$
an  canonical basis of $\R^N$, $H_i=1$ and the coordinates are
$\bx(\bu)=\bu$. This implies that
\begin{align*}
\Omega(\bX,\bxi^*)(\bu)&=A+
\sum_{k=1,\dotsc, N} \be_i\otimes\bxi_i^\t(u_i),\\
\Omega(\bxi,H)(\bu)&=\bc+\sum_{k=1,\dotsc, N} \int_{u_{i,0}}^{u_i}
\dif u_i\, \bxi_i(u_i),
\end{align*}
where $A$ is a constant $N\times M$ matrix and
 $\bc\in\R^N$ is a constant vector, and the orthogonal
conjugate net is given by
\begin{multline*}
\bx(\bu)=\bu-
\bigg[A+\sum_{k=1,\dotsc, N} \be_i\otimes\bxi_i^\t(u_i)\bigg]
\bigg[\sum_{k=1,\dotsc, N}\Omega_i(u_i)\bigg]^{-1} \\
\times\bigg[\bc+
\sum_{k=1,\dotsc, N} \int_{u_{i,0}}^{u_i}
\dif u_i \bxi_i(u_i)\bigg].
\end{multline*}

{\bf 6}. In contrast with the well known Laplace and Levy
transformations there is no  literature on
sequences of  Ribaucour transformations,
however in \cite{demoulin} a permutability theorem was proven
for the 2-dimensional case iterating the Ribaucour transformation twice.
Later in \cite{bianchi}, see also \cite{bianchi1}, it was done in three
dimensional case and in \cite{Eisenhart2} one can find
the extension to any dimension. Recently, in \cite{gt} three 
Ribaucour transformations
were iterated in 3-dimensional space to get some results related with
permutability.  The permutability theorem for the scalar 
fundamental was established in
\cite{jonas}.

\paragraph*{Acknowledgements.} M. M. would like to thank several
conversations with A. Doliwa and P. M. Santini. In particular,
A. Doliwa's historical remarks were quite useful.


\newpage

\begin{thebibliography}{99}


\bibitem{bianchi} L. Bianchi, {\em Rendi. Lincei} {\bf 13} (1904) 361.

\bibitem{bianchi1} L. Bianchi,
{\it Lezioni di Geometria Differenziale}, 
3-a ed., Zanichelli, Bologna (1924).

\bibitem{Darboux1}
G. Darboux,
{\em Le\c{c}ons sur la th\'{e}orie g\'{e}n\'{e}rale des surfaces IV},
 Gauthier-Villars, Paris (1896).
Reprinted by Chelsea Publishing Company,
New York (1972).

\bibitem{Darboux2} G. Darboux,
{\em Le\c{c}ons sur les syst\`{e}mes orthogonaux et les coorden\'{e}es
curvilignes (deuxi\`{e}me \'{e}dition)},
 Gauthier-Villars, Paris (1910) (the first edition was in 1897).
Reprinted by \'Editions Jacques Gabay, Sceaux (1993).


\bibitem{demoulin} M. A. Demoulin, {\em Comp. Rend. Acad. Sci. Paris}
{\bf 150} (1910) 156.

\bibitem{dsm} A. Doliwa, P. M. Santini and  M. Ma\~nas, {\em Transformations
for Quadrilateral Lattices}, to appear (1997).

\bibitem{Eisenhart1}
L. P. Eisenhart, {\it A Treatise on the Differential Geometry of
Curves and Surfaces}, Ginn and Co., Boston (1909).

\bibitem{Eisenhart3}
L. P. Eisenhart, {\em  Trans. Amer. Math. Soc.} {\bf 18} (1917) 111.

\bibitem{Eisenhart2} L. P. Eisenhart, {\em Transformations
of Surfaces}, Princeton University Press, Princeton (1923).
Reprinted by Chelsea Publishing Company, New York (1962).

\bibitem{gt}
E. I. Ganzha and S. P. Tsarev,
{\em On superposition of the auto-B\"acklund transformations for
(2+1)-dimensional integrable systems}, {\tt solv-int/9606003}.

\bibitem{jonas} H. Jonas,  {\em Berl. Math. Ges. Ber. Sitzungsber.} {\bf 14} (1915) 96.

\bibitem{ks} B. G. Konopelchenko and W. K. Schief,
{\em Lam\'e and Zakharov-Manakov systems: Combescure, Darboux and
B\"acklund transformations}, Preprint Applied Mathematics 93/9,
University of New South Wales (1993).

\bibitem{lame} G. Lam\'e, {\em Le\c{c}ons sur la th\'{e}orie des
coorden\'{e}es curvilignes et leurs diverses applications},
Mallet-Bachalier, Paris (1859).

\bibitem{mds} M. Ma\~nas, A. Doliwa and P. M. Santini, {\em Phys. Lett.}
{\bf 232 A}  (1997) 365.


\bibitem{Ribaucour} A. Ribaucour, {\em Comp. Rend. Acad. Sci. Paris}
{\bf 74} (1872) 1489.


\bibitem{schief} W. K. Schief, {\em Inv. Prob.} {\bf 10} (1994) 1185.

\bibitem{tsarev} S. P. Tsarev, {\em Classical differential geometry and integrability
of systems of hydrodynamic type}, in {\em Applications of analytic and
geometrical methods to nonlinear differential equations}, ed.
P. A. Clarkson, Kluwer, Dortrecht (1993).

\bibitem{z} V. E. Zakharov, {\em On Integrability of the Equations Describing
N-Orthogonal Curvilinear Coordinate Systems and Hamiltonian
Integrable Systems of Hydrodynamic Type I: Integration of the
Lam\'e Equations} Preprint (1996).

\bibitem{zm}V. E. Zakharov and S. V. Manakov,
{\em Func. Anal. Appl.} {\bf 19} (1985) 11.
\end{thebibliography}
\end{document}